\documentclass[%
 twocolumn,
superscriptaddress,
preprintnumbers,
 amsmath,amssymb,
 aps,
prl,
]{revtex4-1}

\usepackage{graphics,amssymb,amsmath,epsfig,color}
\usepackage{graphicx}
\usepackage{dcolumn}
\usepackage{bm}
\usepackage{latexsym}
\usepackage{xcolor}
\usepackage[%
  colorlinks=true,
  urlcolor=blue,
  linkcolor=blue,
  citecolor=blue
]{hyperref}

\newcommand{\changed}[1]{{\color{black}#1}}

\begin{document}

\title{Radiative cooling of a superconducting resonator
} 

\author{Mingrui Xu}
\affiliation{Department of Electrical Engineering, Yale University, New Haven, CT 06520, USA}
\author{Xu Han}
\affiliation{Department of Electrical Engineering, Yale University, New Haven, CT 06520, USA}
\author{Chang-Ling Zou}
\affiliation{Department of Electrical Engineering, Yale University, New Haven, CT 06520, USA}
\author{Wei Fu}
\affiliation{Department of Electrical Engineering, Yale University, New Haven, CT 06520, USA}
\author{Yuntao Xu}
\affiliation{Department of Electrical Engineering, Yale University, New Haven, CT 06520, USA}
\author{Changchun Zhong}
\affiliation{Department of Applied Physics, Yale University, New Haven, CT 06520, USA}
\author{Liang Jiang}
\affiliation{Department of Applied Physics, Yale University, New Haven, CT 06520, USA}
\author{Hong X. Tang}
\email{hong.tang@yale.edu}
\affiliation{Department of Electrical Engineering, Yale University, New Haven, CT 06520, USA}

\date{\today}

\begin{abstract}
Cooling microwave resonators to near the quantum ground state, crucial for their operation in the quantum regime, is typically achieved by direct device refrigeration to a few tens of millikelvin. 
However, in quantum experiments that require high operation power such as microwave-to-optics quantum transduction, it is desirable to operate at higher temperatures with non-negligible environmental thermal excitations, where larger cooling power is available.
In this Letter, we present a radiative cooling protocol to prepare a superconducting microwave mode near its quantum ground state in spite of warm environment temperatures for the resonator. 
In this proof-of-concept experiment, the mode occupancy of a 10-GHz superconducting resonator thermally anchored at 1.02~K is reduced to $0.44\pm0.05$ by radiatively coupling to a 70-mK cold load.  
This radiative cooling scheme allows high-operation-power microwave experiments to work in the quantum regime, and opens possibilities for routing microwave quantum states to elevated temperatures. 
\end{abstract}

\maketitle

At thermal equilibrium with a fluctuating environment, a bosonic resonator may possess a finite thermal mode occupancy $\bar{n}_\mathrm{mode}$, which causes decoherence and induces detrimental added noises in performing quantum tasks.
Therefore, refrigeration of microwave devices to ultra-cold temperatures ($T < 100~\mathrm{mK}$) is routinely applied to achieve prolonged coherence 
for quantum operations at microwave frequencies \cite{devoret2013superconducting, xiang2013hybrid}. 
However, cooling through thermal contact faces limits in performing certain tasks. At millikelvin temperatures, the finite cooling power (typically $10^{-5}~\mathrm{W}$) restricts the operation power \cite{hornibrook2015archetecture} and the amount of heat dissipated by the devices
\cite{AmirHeating2014,PulseExcitation2015,midolo2018nano}.
Such limitation hinders the development of, for example, electro-optical quantum state transduction \cite{midolo2018nano,bochmann2013nanomechanical,andrews2014conversion,bagci2014optical,Changling2016,forsch2018groundmicrowave,higginbotham2018harnessing,EOKippenberg2016,rueda2016efficientEO,soltani2017efficientEO,fan2018conversion,nakamura2016bidirectional} and microwave-optical photon entanglement generation \cite{zhong2019heralded} based on hybrid superconducting systems, which are important steps towards the realization of scalable quantum networks \cite{cirac1997quantum,kimble2008quantumInternet}. In such applications, a high-power optical pump is typically required to boost the transduction efficiency or photon entanglement generation rate, 
\changed{while heating due to optical absorption by the dielectrics that host the optical modes become significant \cite{soltani2017efficientEO,AmirHeating2014,PulseExcitation2015,midolo2018nano,ObsorbtionExplain}.}
\par

\begin{figure}[ht!]
\includegraphics{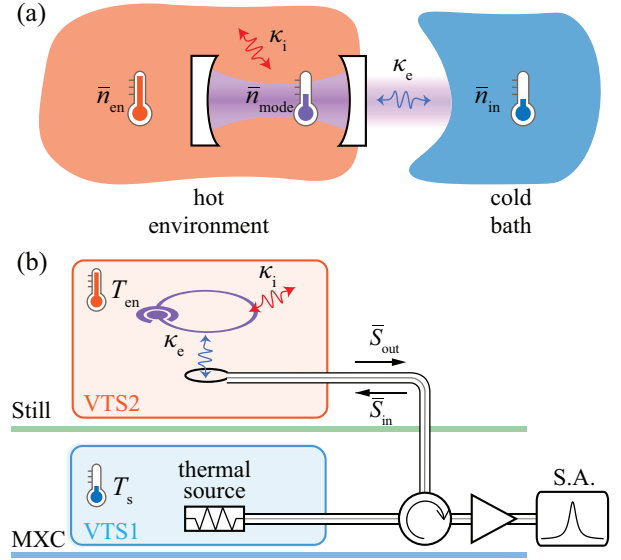}
\caption{\label{fig:diagram} (a) Mechanism of radiative cooling where the superconducting resonator is represented by an optical cavity. $\bar{n}_\mathrm{en}$, $\bar{n}_\mathrm{mode}$ and $\bar{n}_\mathrm{in}$ denote the thermal occupancy of the physical environment, microwave mode and the cold external bath, respectively. $\kappa_\mathrm{i}$ and $\kappa_\mathrm{e}$ are the mode's coupling rate to the environment and external bath, respectively. (b) Schematic of the experimental setup. ``Still'' and ``MXC'' stand for the still plate and the mixing chamber in a dilution refrigerator.
The superconducting resonator (purple) is mounted on a variable temperature stage (VTS2) which provides a variable environment temperature $T_\mathrm{en}$. 
The resistor mounted on the VTS1 serves as a controllable thermal source. 
Output noise from the resonator $\bar{S}_\mathrm{out}$ travels through an output line with amplification before being detected by a spectrum analyzer.
}
\end{figure}

\changed{In this context, selective cooling of microwave resonances \cite{phillips1998nobel,chan2011laser,tan2017quantum}, which relaxes the strict operational temperature requirement, is highly desirable.}
By far the mostly adopted selective cooling approach is the sideband cooling which parametrically couples the target mode to a high-frequency cavity \changed{whose mode is almost at the quantum ground state.} Sideband cooling combined with cryogenic precooling \cite{schliesser2008resolved,park2009resolved,chan2011laser,teufel2011sideband,Safavi2012,peterson2016laser} has been successfully implemented in cavity optomechanics to achieve ground state cooling of mechanical motion. 
However, such a cooling scheme relies on a high cooperativity between the target mode and an optical or a high-frequency microwave mode \cite{aspelmeyer2014cavityOM}, and is yet to be implemented to cool microwave mode residing in a superconducting resonator. 

Alternatively, thermal occupancy of a microwave mode can be selectively suppressed through a direct radiative link to a cold thermal bath, i.e. radiative cooling.
Same radiative cooling principle has been exploited in daytime photonic cooling over the past decade \cite{raman2014sunlightcooling,hossain2016radiative}. 
In typical applications of superconducting resonators, in addition to the inevitable interactions to the environment, each mode is coupled to a bus circuit for input and readout. 
Hence, if one can engineer the bus circuit to have low thermal excitations, at thermal equilibrium the mode temperature will be cooled below the resonator's environment temperature, and even near the quantum ground state.
Similar cooling scheme has recently been proposed to improve detection sensitivity in microwave radiometry applications \cite{Radiometer18,PhotonicReceiver2008}.\par

In this Letter, we report the first experimental implementation of radiative cooling for a superconducting resonator.
In this proof-of-concept demonstration, we create a cold external thermal bath of nearly only vacuum fluctuations (bath occupancy $\bar{n}_\mathrm{in}=0.02^{+0.06}_{-0.02}$) for a superconducting resonator operated at above 1~K. 
A precision noise thermometry calibration employing a Josephson parametric converter
(JPC) amplifier verifies a reduced mode thermal occupancy $\bar{n}_\mathrm{mode}=0.44\pm 0.05$, well below the environment thermal occupancy $\bar{n}_\mathrm{en}=1.56$ at 1.02~K.
\changed{Effectively, the microwave resonance is selectively cooled  through the coupling to the cold external bath.
Therefore, a more effective radiative cooling can be anticipated by further overcoupling the resonator to the cold external bath.}
This selective radiative cooling protocol greatly conserves the limited cooling power of a cryogenic refrigerator at ultra-low temperatures, and relaxes the strict requirement of operational temperature for devices aimed at quantum information processing, such as 
hybrid superconducting electro-optical \cite{EOKippenberg2016,rueda2016efficientEO,soltani2017efficientEO,fan2018conversion} and electro-opto-mechanical \cite{midolo2018nano,bochmann2013nanomechanical,andrews2014conversion,bagci2014optical,Changling2016,forsch2018groundmicrowave,higginbotham2018harnessing,han2016} systems, for which a high optical pump power is routinely required to achieve desired cooperativity between parametrically coupled resonators \cite{aspelmeyer2014cavityOM,EOKippenberg2016,rueda2016efficientEO,soltani2017efficientEO}.


The concept of radiative cooling is schematically illustrated in Fig.~1(a). Immersed in a hot environment (red), a microwave resonator (purple) is radiatively linked to a cold external bath (blue). The thermal occupancy of the microwave mode, the environment, and the external bath are denoted by $\bar{n}_\mathrm{mode}$, $\bar{n}_\mathrm{en}$, and $\bar{n}_\mathrm{in}$, respectively. 
At a given frequency $f$, the average thermal occupancy $\bar{n}$ is linked to its temperature $T$ by the Bose-Einstein distribution $\bar{n}=1/\big(\mathrm{exp}(hf/k_\mathrm{B}T)-1\big)$.
Through thermalization to both the hot environment and cold external bath, with respective coupling rates $\kappa_\mathrm{i}$ and $\kappa_\mathrm{e}$, the mode occupancy can be derived as
\begin{equation}
\label{thermalization}
\bar{n}_\mathrm{mode}=\frac{\kappa_\mathrm{i}\bar{n}_\mathrm{en}+\kappa_\mathrm{e}\bar{n}_\mathrm{in}}{\kappa_\mathrm{i}+\kappa_\mathrm{e}}=\bar{n}_\mathrm{en}-\frac{\kappa_\mathrm{e}}{\kappa}\Delta\bar{n}
\end{equation}
Here, $\kappa=\kappa_\mathrm{i}+\kappa_\mathrm{e}$ is the total decay rate of the mode, and $\Delta\bar{n}=\bar{n}_\mathrm{en}-\bar{n}_\mathrm{in}$ is the thermal occupancy difference between environment and the external bath. 
When the resonator is very overcoupled, e.g. $\kappa_\mathrm{e}/\kappa_\mathrm{i}\gg 1$, the mode temperature is predominately determined by the external bath occupancy $\bar{n}_\mathrm{in}$.\par

A practical implementation of radiative cooling requires construction of a cold external bath and a mode occupancy readout channel to verify the cooling effect, which is depicted in Fig.~\ref{fig:diagram}(b).
In the setup, a variable temperature stage (VTS2) where the resonator (purple) is mounted on provides a tunable environment temperature $T_\mathrm{en}$ with corresponding average thermal occupancy $\bar{n}_\mathrm{en}$.
Through an antenna, the resonator is coupled to a microwave bus circuit which effectively forms the external bath with thermal occupancy $\bar{n}_\mathrm{in}$ to the mode.
In the microwave circuit, a circulator is employed to separate the incoming and outgoing fields to/from the single-sided resonator in order to allow simultaneous radiative cooling and mode occupancy readout. 
Mediated by the circulator, the input to the resonator is terminated with a controllable thermal source, constituted by a impedance-matched resistor thermalized to another variable temperature stage (VTS1) at $T_\mathrm{s}$.
By sweeping VTS1's temperature $T_\mathrm{s}$, thermal photon number $\bar{n}_\mathrm{s}$ of the thermal source output can be varied.
Modeling the transmission link between the thermal source and the resonator as a beam-splitter, with the transmission denoted as $\lambda$, the effective external bath occupancy $\bar{n}_\mathrm{in}$ sensed by the resonator is
\begin{equation}
\label{couplingchannel}
   \bar{n}_\mathrm{in}=\lambda \bar{n}_\mathrm{s}+ (1-\lambda)\bar{n}_\mathrm{eff,link}.
\end{equation}
Here $(1-\lambda)\bar{n}_\mathrm{eff,link}$ is the added noise, determined by the distributed loss and temperature along the physical transmission link whose effective occupancy is denoted by $\bar{n}_\mathrm{eff,link}$. In the ideal case, this added noise approaches zero.
As suggested by Eq.~(\ref{thermalization}), $\bar{n}_\mathrm{in}$ sets the lower bound for the minimal achievable mode occupancy. Therefore, a low-noise radiative cooling channel is crucial to the success of radiative cooling.\par

\begin{figure}[ht]
\includegraphics{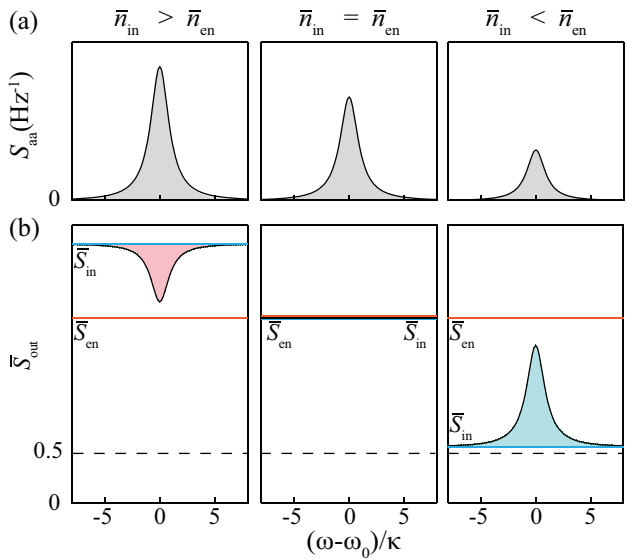}
\caption{\label{expectation} (a) Simulated mode amplitude power spectral density $S_{aa}$ at different external bath temperatures. 
The shaded area in each plot represents the mode occupancy $\bar{n}_\mathrm{mode}$. 
(b) Simulated symmetrized power spectra of output noise $\bar{S}_\mathrm{out}$ under same conditions corresponding to (a). The blue and orange traces indicate symmetrized power spectral density of the external bath $\bar{S}_\mathrm{in}$ and of the environment $\bar{S}_\mathrm{en}$ respectively. A peak emerges on the $\bar{S}_\mathrm{out}$ spectrum when radiative cooling occurs, and inverses when radiative heating happens. 
}
\end{figure}

We first theoretically investigate the radiative cooling effect through the power spectral density $S_{aa}(\omega)$ of intracavity field \cite{supplemental}, where $a$ represents the amplitude of the microwave mode. 
Plotted in Fig.~\ref{expectation}(a) are $S_{aa}(\omega)$ spectra at three different external bath occupancies $\bar{n}_\mathrm{in}$, 
while the environment temperature $T_\mathrm{en}$ is fixed.
In each figure, the mode occupancy $\bar{n}_\mathrm{mode}$ can be extracted by integrating the shaded area, which clearly shows heating or cooling without althering the resonator's environment temperature.

For experimental verification of radiative cooling, we study the output power spectrum from the resonator measured by a classical microwave detector. 
According to the input-output theory \cite{aspelmeyer2014cavityOM}, the output field is comprised of the reflected input and emission from the resonator, thus the resulting symmetrized power spectral density \cite{supplemental} in units of quanta reads 
\begin{equation}
\label{output}
   \bar{S}_\mathrm{out}(\omega)=\mathcal{R}(\omega)\bar{n}_\mathrm{in}+\mathcal{T}(\omega)\bar{n}_\mathrm{en}+\frac{1}{2},
\end{equation}
where $\mathcal{T}(\omega)=\kappa_\mathrm{i}\kappa_\mathrm{e}/\big({(\kappa/2)^2+(\omega-\omega_\mathrm{0})^2}\big)$, $\mathcal{R}(\omega)=1-\mathcal{T}(\omega)$, and $\omega_\mathrm{0}$ is the angular resonant frequency of the resonator. $\mathcal{T}(\omega)$ can be understood as a power transmission spectrum from the environment to the microwave circuit through the resonator, which is a localized Lorenztian-shaped function peaked at the resonant frequency.
Complementary to $\mathcal{T}(\omega)$, 
$\mathcal{R}(\omega)$ is the power reflection spectrum of a single-sided resonator. 
Denoting the incident noise power spectral density from the external bath as $\bar{S}_\mathrm{in}=\bar{n}_\mathrm{in}+\frac{1}{2}$, Eq.~(2) can be simplified as 
\begin{equation}
\label{delta}
   \bar{S}_\mathrm{out}(\omega)=\bar{S}_\mathrm{in}+\mathcal{T}(\omega)\Delta\bar{n}.
\end{equation}
In the following discussion we define $\Delta\bar{S}_\mathrm{out}(\omega)$ as $\Delta\bar{S}_\mathrm{out}(\omega)=\bar{S}_\mathrm{out}(\omega)-\bar{S}_\mathrm{in}$ to denote the non-flat spectral feature in $\bar{S}_\mathrm{out}(\omega)$.\par

By sweeping the thermal source temperature $T_\mathrm{s}$, thereby effectively changing $\bar{n}_\mathrm{in}$, the output power spectral density $\bar{S}_\mathrm{out}$ varies accordingly, as shown in Fig.~\ref{expectation}(b). The blue and orange traces mark the level of $\bar{S}_\mathrm{en}$ and $\bar{S}_\mathrm{in}$ respectively as references, where $\bar{S}_\mathrm{en}=\bar{n}_\mathrm{en}+\frac{1}{2}$.
When $\bar{n}_\mathrm{in}=\bar{n}_\mathrm{en}$, meaning the resonator as well as the microwave circuit are both in thermal equilibrium at the same temperature, from Eq.~(\ref{delta}) we arrive at $\bar{S}_\mathrm{out}(\omega)=\bar{S}_\mathrm{in}$, hence $\bar{S}_\mathrm{out}(\omega)$ in the middle panel of Fig.~2(b) is a flat trace.
For $\bar{n}_\mathrm{in} < \bar{n}_\mathrm{en}$, $\Delta\bar{S}_\mathrm{out}(\omega)>0$, leading to a positive Lorentzian shaped peak on a flat background for $\bar{S}_\mathrm{out}(\omega)$, which is the key signature of radiative cooling. 
\changed{Contrarily,} when $\bar{n}_\mathrm{in} > \bar{n}_\mathrm{en}$, $\Delta\bar{S}_\mathrm{out}(\omega)<0$, a Lorentzian shaped dip emerges in the spectrum as in the left panel of Fig.~2(b), which is indicative of radiative heating. 
Since $\Delta\bar{S}_\mathrm{out}(\omega)$ scales with $\Delta \bar{n}$, it reveals how much colder the external bath is compared to the resonator's environment.
\par

In the experiment, the mode occupancy $\bar{n}_\mathrm{mode}$ can be unambiguously calibrated through a set of measurements: 
First, $\mathcal{R}(\omega)$ can be characterized by probing the resonator with a weak coherent tone \cite{supplemental}. 
Subsequently, $\kappa_\mathrm{e}$, $\kappa_\mathrm{i}$ can be deduced from a theoretical fit of $\mathcal{R}(\omega)$.
Second, the resonator's environment thermal occupancy $\bar{n}_\mathrm{en}$ can be obtained through the readout of the resonator's physical temperature by a temperature sensor. 
Third, with a calibrated output line, one can experimentally obtain $\Delta\bar{S}_\mathrm{out}(\omega)$ by measuring $\bar{S}_\mathrm{out}(\omega)$ and $\bar{S}_\mathrm{out,off}(\omega)$, where $\bar{S}_\mathrm{out,off}(\omega)$ is the output noise spectrum measured when the resonance is externally tuned far away from the measurement frequency window. 
From Eq.~(\ref{delta}), we arrive at $\bar{S}_\mathrm{out,off}(\omega)=\bar{S}_\mathrm{in}$ under the condition of $|\omega-\omega_\mathrm{0}|\gg\kappa$. 
Hence $\Delta\bar{S}_\mathrm{out}(\omega)$ can be found as $\Delta\bar{S}_\mathrm{out}(\omega)=\bar{S}_\mathrm{out}(\omega) -\bar{S}_\mathrm{out,off}(\omega)$, from which $\Delta \bar{n}$ can be extracted.
Finally, with the knowledge of $\kappa_\mathrm{i}$, $\kappa_\mathrm{e}$, $\bar{n}_\mathrm{en}$, and $\Delta \bar{n}$, through Eq.~(\ref{thermalization}), $\bar{n}_\mathrm{mode}$ can be readily deduced.\par


The above analysis suggests that the scheme illustrated in Fig.~1(b) is capable of performing radiative cooling as well as calibrating the mode occupancy. Therefore, such a setup was experimentally implemented. 
\changed{The superconducting resonator of interest possesses a resonant frequency of $\omega_\mathrm{0}=2\pi \times 10.53~\mathrm{GHz}$,} which can be precisely tuned by a DC external magnetic field through a persistent current-induced-modification of kinetic inductance \cite{2018TunableOuroboros}. 
Characterized by a VNA with a weak coherent tone which excites about 3 photons in the resonator, the intrinsic and external coupling rates are found to be $\kappa_\mathrm{i}=2\pi \times (113\pm 1)~\mathrm{kHz}$ and $\kappa_\mathrm{e}=2\pi \times (298 \pm 2)~\mathrm{kHz}$, respectively. 
VTS1 and VTS2 are mounted on the mixing chamber and still plate respectively in a dilution refrigerator.
Between components, superconducting NbTi coaxial cables are used to provide low-loss links for radiative heat exchange.
On the output line, a low-noise pre-amplifier\textemdash Josephson parametric converter (JPC) \cite{bergeal2010JPC}\textemdash is implemented followed by a high-electron-mobility transistor (HEMT) amplifier to provide precise detection of the output noise power from the resonator.


\begin{figure}[h]
\includegraphics{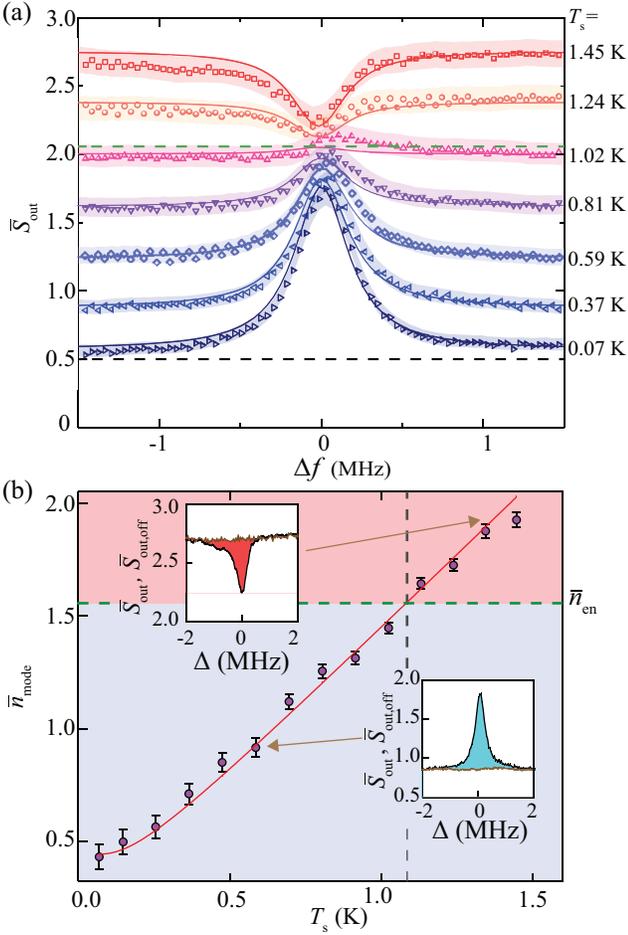}
\caption{\label{fig3} (a) Calibrated $ \bar{S}_\mathrm{out}$ spectra measured at different $T_\mathrm{s}$. Shaded areas indicate the error bands. Solid traces represent the theoretical expectations independently calculated. During the measurement, $T_\mathrm{en}$ is fixed at 1.02~K, where $\bar{S}_\mathrm{en}$ is represented by the green dashed trace. (b) Deduced mode occupancy $\bar{n}_\mathrm{mode}$ as a function of the thermal source temperature $T_\mathrm{s}$. The red solid curve represents the theoretical prediction. The horizontal dashed line marks $\bar{n}_\mathrm{en}$ as a reference. The vertical dashed line indicates the transition from radiative cooling to heating. In each inset, the black trace represents the $\bar{S}_\mathrm{out}(\omega)$ spectrum and the brown trace represents $\bar{S}_\mathrm{out,off}(\omega)$ spectrum, which difference (shaded) is $\Delta \bar{S}_\mathrm{out}(\omega)$. }
\end{figure}

To enable absolute measurement of the output noise spectrum $\bar{S}_\mathrm{out}(\omega)$, the system is precisely calibrated. 
The output line calibration, i.e. identifying the gain $G_\mathrm{0}$ and the added noise that refers to the output of the resonator, is made possible by noise thermometry calibration while varying the resonator's environment temperature $T_\mathrm{en}$ by VTS2, where the temperature sensor's readout of $T_\mathrm{en}$ is used as the reference \cite{supplemental}.
Similar thermometry calibration of the output line gain $G_\mathrm{s}$ referred back to the output of the thermal source was conducted by varying the temperature $T_\mathrm{s}$ through heating VTS1. Comparing $G_\mathrm{0}$ and $G_\mathrm{s}$, the attenuation of the transmission link between the thermal source on VTS1 and the resonator can be characterized, which reveals $\lambda=G_\mathrm{s}/G_\mathrm{0}=91\% \pm 4\%$. Moreover, added noise of the transmission link is found to be $(1-\lambda)\bar{n}_\mathrm{eff, link}=0.02^{+ 0.06}_{-0.02}$ \cite{supplemental}.\par

To demonstrate radiative cooling or heating, $\bar{S}_\mathrm{out}(\omega)$ is measured when the external bath temperature is swept across $T_\mathrm{en}$.
Figure~3(a) plots a set of calibrated output noise spectra $\bar{S}_\mathrm{out}(\omega)$ at different thermal source temperatures $T_\mathrm{s}$ ranging from 70~mK to 1.45~K by heating the VTS1. 
Throughout this set of measurements, the resonator's environment temperature $T_\mathrm{en}$ is fixed at 1.02~K, providing a constant environment thermal occupancy $\bar{n}_\mathrm{en}=1.56$.
Measurement errors limited by the calibration sensitivity are represented by shaded areas.
As the temperature $T_\mathrm{s}$ increases, along with the rising background associated with $\bar{S}_\mathrm{in}$, we observed a peak or dip on each $\bar{S}_\mathrm{out}(\omega)$ spectrum, indicative of radiative cooling or heating.
In excellent agreement with the data, theoretical predictions (solid traces) are each independently calculated based on Eq.~(\ref{delta}), $T_\mathrm{en}$ read out by the temperature sensor and $\bar{S}_\mathrm{in}$ extracted from measured $\bar{S}_\mathrm{out,off}(\omega)$ when the resonance is far detuned by more than 30 linewidths. 
Slight distortion from an ideal Lorenztian lineshape is observed in measured $\bar{S}_\mathrm{out}(\omega)$ spectra. 
It could be attributed to the interference between the output noise from resonator and the feed-through noise due to the imperfect isolation of the circulator, which effectively results in an asymmetric Fano resonance \cite{tu2010fano,li2011fano,supplemental}.\par

At each thermal source temperature $T_\mathrm{s}$, based on measured output noise spectra, the mode occupancy $\bar{n}_\mathrm{mode}$ can be deduced as discussed earlier.
Here in each inset of Fig.~3(b) plots measured $\bar{S}_\mathrm{out}(\omega)$ and $\bar{S}_\mathrm{out,off}(\omega)$ spectra, where $\Delta \bar{S}_\mathrm{out}$ is found as the difference. 
Exploiting the fact that the hight of $\Delta \bar{S}_\mathrm{out}$ is proportional to $\Delta \bar{n}$, \changed{we extract $\Delta \bar{n}$ as $\Delta \bar{n}=\big( \kappa/(2\pi\kappa_\mathrm{e}\kappa_\mathrm{i})\big) \int^\infty_{-\infty}\Delta \bar{S}_\mathrm{out} \mathrm{d}\omega.$}
Combining with obtained $\bar{n}_\mathrm{en}$ and coupling rates, the mode occupancy $\bar{n}_\mathrm{mode}$ can be deduced from Eq.~(\ref{thermalization}), which results
are plotted in Fig.~3(b) as a function of $T_\mathrm{s}$. 
Theoretical prediction based on Eqs.~(\ref{thermalization}), (\ref{couplingchannel}), $\bar{n}_\mathrm{en}$ and independently performed transmission link calibration is shown as the solid red trace, which is in excellent agreement with the measurement results.
Marked by the vertical dashed line is the transition from radiative cooling to radiative heating, where the thermal source temperature is at $T_\mathrm{s}=1.08~\mathrm{K}$. This transition temperature is slightly higher than $T_\mathrm{en}=1.02~\mathrm{K}$, due to the imperfect transmission link between the controllable thermal source and the resonator. 
The lowest mode occupancy $\bar{n}_\mathrm{mode}=0.44\pm 0.05$ is observed when the thermal source is at its base temperature $T_\mathrm{s}=70~\mathrm{mK}$ ($\bar{n}_\mathrm{s}= 0.001$). 
In such a condition, according to Eq.~(\ref{couplingchannel}), the external bath has almost zero thermal occupancy $\bar{n}_\mathrm{in}\approx (1-\lambda)\bar{n}_\mathrm{eff,link}=0.02^{+ 0.06}_{-0.02}$. 
Hence Eq.~(\ref{thermalization}) can be reduced to $\bar{n}_\mathrm{mode}\approx\frac{\kappa_\mathrm{i}}{\kappa}\bar{n}_\mathrm{en}$, 
where the cooling effectiveness is determined by the coupling condition of the resonator. \par

\changed{To achieve further suppression of the superconducting mode occupancy, one can configure the resonator to be more overcoupled to the external cold bath.
It is therefore desirable to have a high-intrinsic-\emph{Q} resonance to start with, so that the over-coupling condition can be achieved with a reasonable total decay rate.}
While in the present experiment, $\kappa_\mathrm{e}$ is chosen to be comparable to $\kappa_\mathrm{i}$ ($\kappa_\mathrm{e}$/$\kappa_\mathrm{i}=2.6$) for demonstration purpose and a relatively high noise thermometry calibration sensitivity. 
Because otherwise, under the condition of $\kappa_\mathrm{e}$/$\kappa_\mathrm{i} \gg 1$, the maximum value of $\mathcal{T}(\omega)$ will reduce to close to zero, leading to an almost flat $\bar{S}_\mathrm{out}(\omega)$ spectrum.
\changed{Consequently, very limited} portion of the environment thermal radiation will be coupled into the transmission line, \changed{which will degrade} the sensitivity of the output line calibration using the resonator's environment temperature $T_\mathrm{en}$ as a reference. 
Nonetheless, in the current setup, by moving the antenna closer to the superconducting resonator, a higher external coupling rate $\kappa_\mathrm{e}=5~\mathrm{MHz}$ ($\kappa_\mathrm{e}$/$\kappa_\mathrm{i}=44$) can be obtained. Given otherwise identical experimental parameters, one can expect the mode occupancy to be suppressed to $\bar{n}_\mathrm{mode}=0.05$ from $\bar{n}_\mathrm{en}=1.52$ at $T_\mathrm{en}=1~\mathrm{K}$, which is closer to the ground state.
To improve noise thermometry calibration sensitivity under extremely overcoupled condition, one can exploit degenerate parametric amplifier \cite{optimal2018Lehnerd} which is free from the $1/2$ quantum noise or a qubit as a noise power detector  \cite{Zhixin2019Radiometer}.
\changed{Looking forward, radiative cooling can be applied to cool superconducting resonators physically anchored at even higher temperatures to close to quantum ground state. For example, for an ultra-high-\emph{Q} 10-GHz superconducting resonator physically immersed in liquid helium (4.2~K) \cite{kuhr2007ultrahigh,padamsee2001science}, by preparing it to be extremely overcoupled ($\kappa_\mathrm{e}$/$\kappa_\mathrm{i}=100$) to a millikelvin cold bath, the mode occupancy can be reduced to $n_\mathrm{mode}=0.1$.}
\par

\changed{To conclude, we demonstrate a cooling protocol for superconducting resonators through a radiative link to a cold bath, and a readout scheme to precisely calibrate the mode occupancy.}
With insignificant experimental overhead, radiative cooling enables routing microwave photons\cite{Xiang2017} with reduced added noises at relaxed refrigeration conditions where higher cooling power is available. This result marks an important pathway for the realization of quantum interfaces between superconducting and photonic circuits.
Immediate applications of radiative cooling for superconducting resonators include quantum state transduction \cite{midolo2018nano,bochmann2013nanomechanical,andrews2014conversion,bagci2014optical,Changling2016,forsch2018groundmicrowave,higginbotham2018harnessing,EOKippenberg2016,rueda2016efficientEO,soltani2017efficientEO,fan2018conversion,nakamura2016bidirectional} and entangled photon pair generation \cite{zhong2019heralded}.
Moreover, the cooling approach can also be generalized to selectively suppress thermal excitations in other bosonic excitations, such as phonons \cite{Radiometer18,fu2019phononic} and magnons \cite{Xufeng2014MagnonCouple,Nakamura2014MagnonCouple}.\par

\begin{acknowledgments}
The authors would like to thank M. H. Devoret, S. Shankar, Z. Wang for many inspirational discussions and their immense help on instrumentation. The authors thank A. A. Sayem, A. W. Bruch and J. B. Surya for critical reading of the manuscript and M. Power, J. Agresta, C. Tillinghast, and M. Rooks for assistance in device fabrication. This work is supported by ARO grant W911NF-18-1-0020. The authors also acknowledge partial supports from AFOSR MURI (FA9550-14-1-0052, FA9550-15-1-0015), DOE (DE-SC0019406), NSF (EFMA-1640959) and the Packard Foundation. 
\end{acknowledgments}

\appendix

\bibliographystyle{apsrev4-1}

\end{document}